\documentclass[twocolumn,prb,showpacs,superscriptaddress,longbibliography]{revtex4}
\usepackage{amssymb}
\usepackage{amsmath}
\usepackage{bbm}
\usepackage{graphicx}
\usepackage[colorlinks=true]{hyperref}

\usepackage{tikz}
\usetikzlibrary{positioning,arrows,patterns}
\usetikzlibrary{decorations.markings,decorations.pathmorphing}
\pgfmathtruncatemacro\distance{1}
\tikzset{
  normal/.style={draw=blue, very thick},
  photon/.style={draw=red, very thick, decorate, decoration={snake}},
  fermion/.style={draw=blue, very thick, postaction={decorate}, decoration={markings, mark=at position .55 with {\arrow{stealth}}}},
  vertex/.style={draw=blue, very thick, shape=circle, fill=blue, minimum size=3pt, inner sep=0pt},
}

\pgfrealjobname{diagapprox}

\newcommand{\dd}{\mathrm{d}}
\renewcommand{\vec}[1]{\boldsymbol{#1}}

\begin{document}

\author{Jan Gukelberger}
\email[Corresponding author: ]{gukelberger@phys.ethz.ch} 
\affiliation{Theoretical Physics, ETH Zurich, 8093 Zurich, Switzerland}
\author{Li Huang}
\affiliation{Department of Physics, University of Fribourg, 1700 Fribourg, Switzerland} 
\author{Philipp Werner}
\affiliation{Department of Physics, University of Fribourg, 1700 Fribourg, Switzerland}
\title{On the dangers of partial diagrammatic summations: Benchmarks for the two-dimensional Hubbard model in the weak-coupling regime}

\date{\today}

\hyphenation{}

\begin{abstract}
We study the two-dimensional Hubbard model in the weak-coupling regime and compare the self-energy obtained from various approximate diagrammatic schemes to the result of diagrammatic Monte Carlo simulations, which sum up all weak-coupling diagrams up to a given order. While dynamical mean-field theory provides a good approximation for the local part of the self-energy, including its frequency dependence, the partial summation of bubble and/or ladder diagrams typically yields worse results than second order perturbation theory. Even widely used self-consistent schemes such as $GW$ or the fluctuation-exchange approximation (FLEX) are found to be unreliable. Combining the dynamical mean-field self-energy with the nonlocal component of $GW$ in $GW$+DMFT yields improved results for the local self-energy and nonlocal self-energies of the correct order of magnitude, but here, too, a more reliable scheme is obtained by restricting the nonlocal contribution to the second order diagram. FLEX+DMFT is found to give accurate results in the low-density regime, but even worse results than FLEX near half-filling. 
\end{abstract}

\pacs{71.10.Fd}

\maketitle
\section{Introduction}

Numerically exact approaches for the solution of correlated lattice models such as the Hubbard model are limited to one dimension,\cite{white1992,schollwock2011} small lattices,\cite{White1989} weak coupling,\cite{prokofev2007bdm} high temperature,\cite{Fuchs2011, LeBlanc2013} or to models with particular symmetries and fillings. It is therefore important to develop approximate methods which work in the thermodynamic limit, in more than one dimension, and in the most interesting range of interactions and densities. Typically this means interactions comparable to the bandwidth and densities  close to but not at half band filling. One widely used scheme is the dynamical mean-field theory (DMFT),\cite{Georges1996} which corresponds to the summation of all local self-energy diagrams, via a self-consistent impurity construction. This approximation becomes exact in the limit of infinite dimensions,\cite{Metzner1989,Mueller-Hartmann1989} as well as in the atomic limit and the noninteracting limit. It also captures strong-correlation phenomena such as the Mott transition. The DMFT approximation however neglects spatial fluctuations and thus cannot be expected to capture all the relevant physics in low-dimensional systems. One possibility is to extend DMFT into a cluster-DMFT formalism,\cite{Maier2005} which explicitly treats the correlations within some small cluster. Another possibility is to implement a diagrammatic expansion around the DMFT solution by computing the impurity vertex.\cite{Toschi2007,Rubtsov2008, Rohringer2013, Taranto2014} Both approaches are computationally expensive and hence 
in practice are limited to small clusters, or involve the truncation of the diagrammatic expansion to leading order terms, or some ladder-type series. Especially in view of possible applications to realistic multi-band systems, it is thus desirable to devise simpler, computationally less demanding schemes. 

One strategy, which has been recently explored in simple model contexts,\cite{Ayral2012,Ayral2013,Hansmann2013} is to combine the local DMFT self-energy with the nonlocal component of some many-body perturbation theory (MBPT) such as second-order weak-coupling perturbation theory ($\Sigma^{(2)}$) or the $GW$\cite{hedin1965} approximation.\cite{Sun2002,Biermann2003} Alternative schemes, such as the combination with the nonlocal self-energy from the fluctuation-exchange approximation (FLEX)\cite{Bickers1989,Bickers1989flex1} or the $T$-matrix approximation (TMA),\cite{Galitzkii1958} will also be considered in this work.\footnote{We will not test more sophisticated diagrammatic methods, which require the calculation and manipulation of the frequency-dependent impurity vertex, such as the parquet approximation.}  
The advantage of an approach which combines DMFT and MBPT at the single-particle (self-energy) level is that the computational effort is comparable to single-site DMFT and that the extension to multi-band systems is rather straightforward. The hope is that the local self-energy contribution from DMFT captures the strong-correlation effects while approximately correct nonlocal components are introduced by  the weak-coupling approach. 

In a sufficiently weakly correlated system, the local DMFT contribution may not be needed, so that self-consistent re-summations of certain classes of weak-coupling diagrams, such as bubble and/or ladder diagrams, provide an adequate description. While some tests of the 
$GW$,\cite{pollehn1998aga,schindlmayr1998ste} 
TMA\cite{schindlmayr1998ste,Friesen2010} or FLEX\cite{Bulut1993} approaches have been published, we still lack a clear picture about the importance of the different diagram classes, and the beneficial or detrimental effect of self-consistent partial re-summations. 

The purpose of this study is to shed some light on these issues by benchmarking the approximate self-energies obtained from various MBPT schemes, DMFT and combined MBPT+DMFT approaches against results obtained in diagrammatic Monte Carlo (DiagMC) calculations, which take into account all diagrams up to a certain order. More specifically, we focus on the single-band Hubbard model on the square lattice
\begin{equation}
H=\sum_{i\ne j,\sigma} t_{ij} c^\dagger_{i\sigma}c_{j\sigma}+\sum_i [Un_{i\uparrow}n_{i\downarrow}-\mu (n_{i\uparrow}+n_{i\downarrow})],
\label{hamilt}
\end{equation}
with $t_{ij}=t$ for $i$ and $j$ nearest-neighbor lattice sites, and zero otherwise. The Fourier transform of the hopping matrix is hence $\epsilon_{\vec{k}}=-2t(\cos k_x+\cos k_y)$. We choose the hopping amplitude $t=1$ as the unit of energy. Our test calculations will be limited to the weak-coupling regime $U\lesssim 4t$ (half bandwidth), because in this regime converged DiagMC data can be obtained. Such a comparison is useful despite this limitation, since a controlled approximation based on weak-coupling diagrams, or a combination of weak-coupling diagrams and DMFT, should behave properly in this limit.   

The paper is organized as follows. In Section~\ref{methods}, we briefly discuss a number of established approximations (DMFT, $\Sigma^{(2)}$, $GW$, TMA, FLEX) and the DiagMC method. In Section~\ref{Results}, we benchmark the quality of the local DMFT self-energy, the local and nonlocal MBPT self-energies and various MBPT+DMFT approaches. We also study the convergence properties of partial summations of different classes of weak-coupling diagrams. Section~\ref{Conclusions} contains a summary and conclusion.

\section{Methods}\label{methods}

\subsection{Dynamical mean-field theory}

%Dynamical mean-field theory\cite{Georges1996} maps a lattice model onto a self-consistently defined quantum impurity model, which assumes a momentum-independent self energy $\Sigma(\vec{k},i\omega_n)=\Sigma_\text{DMFT}(i\omega_n)$. The self-consistency condition demands that the impurity Green's function is identical to the local lattice Green's function: $\int (\dd k) \, G(\vec{k},i\omega_n)=G_\text{imp}(i\omega_n)$, where $\int (\dd k)$ denotes a normalized integral over the first Brillouin zone. This fixes the noninteracting impurity Green's function $\mathcal{G}_0(i\omega_n)$, which plays the role of the dynamical mean field. In practice, the self-consistent solution is found by iterating the following steps:

Dynamical mean-field theory\cite{Georges1996} maps a lattice model onto a self-consistently defined quantum impurity model described by the action
\begin{align}
S_\text{DMFT}=&\int_0^{1/T} d\tau [Un_{i\uparrow}(\tau)n_{i\downarrow}(\tau)-\mu (n_{i\uparrow}(\tau)+n_{i\downarrow}(\tau))]\nonumber\\
&+\sum_\sigma\int_0^{1/T} d\tau d\tau' c_\sigma^\dagger(\tau)\Delta_\sigma(\tau-\tau')c_\sigma(\tau'),
\end{align}
where $T$ is the temperature and $\Delta(\tau)$ is the hybridization function. 
%The approximation which enables this mapping is the assumption of a momentum-independent self energy 
In this approximation the self-energy is assumed to be momentum-independent, i.e., $\Sigma(\vec{k},i\omega_n)=\Sigma_\text{DMFT}(i\omega_n)$. The DMFT self-consistency condition demands that the impurity Green's function is identical to the local lattice Green's function: $\int (\dd k) \, G(\vec{k},i\omega_n)=G_\text{imp}(i\omega_n)$, where $\int (\dd k)$ denotes a normalized integral over the first Brillouin zone. This condition fixes the noninteracting impurity Green's function $\mathcal{G}_0(i\omega_n)$, or equivalently the hybridization function $\Delta(i\omega_n)=i\omega_n+\mu-1/\mathcal{G}_0(i\omega_n)$, which plays the role of the dynamical mean field. In practice, the self-consistent solution is found by iterating the following steps (here formulated in terms of the ``mean field" $\mathcal{G}_0$):
\begin{enumerate}
\item Solve impurity model: given $\mathcal{G}_0(i \omega_n)$, compute $G_\text{imp}(i \omega_n)$,
\item Extract self-energy $\Sigma_\text{DMFT}(i \omega_n)=\mathcal{G}_0^{-1}(i \omega_n)-G_\text{imp}^{-1}(i \omega_n)$,
\item DMFT approximation: $\Sigma(\vec{k},i \omega_n)=\Sigma_\text{DMFT}(i \omega_n)$,
\item Compute $G_{\text{loc}}(i \omega_n)=\int (\dd k) [i\omega_n+\mu-\epsilon_{\vec{k}}-\Sigma_\text{DMFT}(i \omega_n)]^{-1}$,
\item DMFT self-consistency: $\mathcal{G}_0^{-1}(i \omega_n)=\Sigma_\text{DMFT}(i \omega_n)+G_{\text{loc}}^{-1}(i \omega_n)$.
\end{enumerate} 
%
%For brevity, the frequency argument $i \omega_n$  of the Green's functions and self-energies has been suppressed. 
%Our DMFT results are calculated using a strong-coupling continuous-time impurity solver\cite{Werner2006} and are thus numerically exact within statistical errors. 
In the present study, the impurity models are solved using a strong-coupling continuous-time quantum Monte Carlo impurity solver\cite{Werner2006} which is numerically exact within statistical errors.

The DMFT self-energy corresponds to the sum of all one-particle irreducible self-energy diagrams which contain only local dressed propagators $G_{\text{loc}}$.\footnote{Here, the local Green's function is computed using the DMFT approximation, and is in general not identical to the exact local lattice Green's function.} %\cite{footnote_DMFT} 
This approximation becomes exact in the limit of infinite dimensions.\cite{Metzner1989, Mueller-Hartmann1989} In low-dimensional systems it is \emph{a priori} unclear how important the neglected contributions from diagrams with nonlocal propagators are, even for the local self-energy.

\begin{figure}
    \includegraphics[width=\columnwidth]{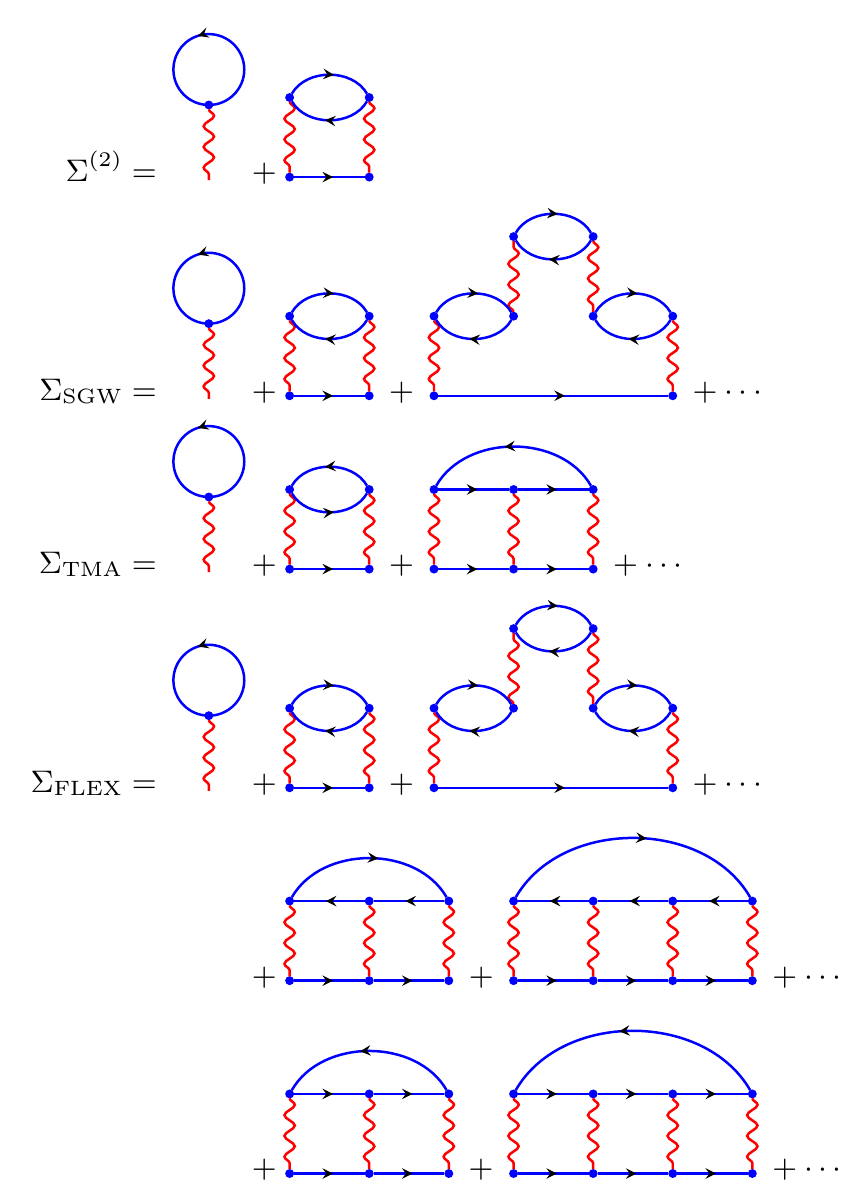}
    \caption{(Color online) Illustration of different many-body approximations to the self-energy. The red wiggly lines represent the on-site interaction $U$. The blue lines with arrows correspond to either bare propagators $G_0$, or (in the case of self-consistent perturbation theory) bold propagators $G$. The first (``tadpole'') diagram is the Hartree term. $\Sigma^{(2)}$ is the second order perturbation theory. $\Sigma_\text{SGW}$ is the spin-dependent $GW$ approximation. The spin-independent $GW$ approximation\cite{hedin1965} in addition contains all the bubble diagrams with an odd number of interaction lines (not shown). $\Sigma_\text{TMA}$ is the  $T$-matrix approximation.\cite{Galitzkii1958} Finally, $\Sigma_\text{FLEX}$ is the fluctuation exchange approximation.\cite{Bickers1989,Bickers1989flex1} \label{fig:mbpt}}
\end{figure}

\subsection{Weak-coupling approaches} \label{sec:weak}

MBPT encompasses 
several techniques which, motivated by diagrammatic perturbation expansions, approximate the electron self-energy $\Sigma$ at different levels. Methods like $GW$ or FLEX are frequently considered since they can treat spatial fluctuations, are easily implemented and appealing on physical grounds. While the truncation of the weak-coupling series for the self-energy at the first order yields the Hartree-Fock approximation, which for the paramagnetic Hubbard model just amounts to a mean-field shift of the chemical potential, the second-order approximation, displayed in the top row of Fig.~\ref{fig:mbpt}, includes some nontrivial correlation effects and creates a nontrivial frequency and momentum dependence in the self-energy. We will denote the second order approximation with bare propagators by $\Sigma^{(2)}$.

Because a systematic computation of the weak-coupling expansion to high orders requires rather involved and costly numerical computations (see Section~\ref{diagmc}), and has only recently become feasible, typical approaches to go beyond second order single out specific diagram topologies, which may be expected to be dominant in some scenarios, and sum these diagrams analytically to infinite order by means of Dyson-like equations. In addition to the choice of topologies to be included, the diagrams can be evaluated with bare propagators $G_0$ or interacting propagators $G$. If the self-energy is derived from a functional of the self-consistently computed $G$, the approximation can be shown to satisfy certain conservation laws.\cite{Baym1962} In practice, however, bare expansions in terms of $G_0$ are often found to be more reliable.\cite{Tsuji2013,Kozik2014} (MBPT approaches which involve both bare and interacting propagators have also been proposed,\cite{Janko1997} but will not be considered here.)

A further issue is whether the on-site interaction is considered to act only between different spin species or also between identical spins. While the full diagrammatic theory respects the Pauli exclusion principle by construction and is hence oblivious to this choice, diagrammatic approximations often violate this constraint. For instance, the two diagrams in the first row of Fig.~\ref{fig:mbpt} constitute all terms of the weak-coupling expansion up to third-order corrections in $U$ if the interaction only acts between different spins. With a spin-independent interaction, however, both diagrams would come with a factor of two from the spin sum associated with the fermion loops; this factor would need to be compensated by a first- and a second-order exchange diagram with the same value but opposite sign as the shown diagrams. At higher orders many more diagrams need to be included to fully compensate terms violating the exclusion principle. We therefore adopt the spin-dependent formalism for all the approximations shown in Fig.~\ref{fig:mbpt}.

In the $GW$ approximation,\cite{hedin1965,aulbur1999,Aryasetiawan1998} the self-energy is given by the product of the Green's function $G$ and the screened Coulomb interaction $W$, where only contributions from the bubble diagrams are considered in the calculation of $W$. 
In scenarios with long-range Coulomb interactions the individual diagrams with bubble insertions are strongly divergent and it is hence essential to sum the infinite series into a screened interaction.
We consider both the self-consistent $GW$ scheme, where all propagator lines denote the dressed $G$, and the ``one-shot" approach $G_0 W_0$, where the diagrams are evaluated with bare propagators. While most $GW$ calculations assume a spin-independent interaction, this leads to the inclusion of $W$ diagrams with an odd number of bare interaction lines, which vanish for the on-site interaction of model (\ref{hamilt}). As this choice (and the neglect of diagrams restoring the Pauli principle) effectively removes spin-fluctuations, which can be expected to be relevant particularly in the vicinity of half filling, we also consider the spin-dependent $GW$ approximation (S$GW$), which retains only the even-order diagrams, as illustrated in the second row of Fig.~\ref{fig:mbpt}. The TMA approach, on the other hand, sums the series of particle-particle ladder diagrams (third row of Fig.~\ref{fig:mbpt}), which dominate the diagrammatic series when the typical inter-particle distance is much larger than the range of the interaction.\cite{Galitzkii1958} In the FLEX approach,\cite{Bickers1989,Bickers1989flex1} finally, bubble, particle-particle and particle-hole ladder diagrams are included (bottom part of Fig.~\ref{fig:mbpt}), which means that this approximation  treats the interaction of electrons via spin, density and pairing fluctuations on equal footing. All of these approximations 
have been widely used to study the properties of interacting lattice models or realistic materials in the weak-to-intermediate correlation regime.\cite{Aryasetiawan1998,Onida2002,Sakakibara2014,Onari2014} 

The computational steps for the spin-independent $GW$ approximation are as follows:
\begin{enumerate}
\item Initialize the self-energy $\Sigma_{GW}(\vec{k},i\omega_n)=0$.
\item Calculate the Green's function $G(\vec{k},i\omega_n) = 1 / [i\omega_n + \mu - \epsilon_{\vec{k}} - \Sigma_{GW}(\vec{k},i\omega_n)]$.
\item Calculate the particle-hole polarization function $ \Pi_{GW}(\vec{k},i\nu_m)  = 2 (T/N_k) \sum_{\vec{q}} \sum_{i\omega_n}  G(\vec{q},i\omega_n) G(\vec{q}-\vec{k},i\omega_n - i\nu_m)$.
\item Calculate the fully screened interaction 
$W(\vec{k},i\nu_m) = 1 / [v^{-1}_{\vec{k}} - \Pi_{GW}(\vec{k},i\nu_m)]$. 
For the Hubbard model \eqref{hamilt}, the bare interaction is $v_{\vec{k}} = U$. 
\item Calculate 
the new self-energy
$\Sigma_{GW}(\vec{k},i\omega_n) = -(T/N_k) \sum_{\vec{q}} \sum_{i\nu_m} G(\vec{q},i\omega_n - i\nu_m) W(\vec{k}-\vec{q},i\nu_m)$.
\item Go to step 2 until converged results for $\Sigma_{GW}(\vec{k},i\omega_n)$ and $\Pi_{GW}(\vec{k},i\nu_m)$ are obtained.
\end{enumerate}
Here, $\omega_n$ denotes a fermionic Matsubara frequency and $\nu_m$ a bosonic Matsubara frequency. 
$N_k$ is the number of momentum points in the discretized Brillouin zone.
Note that in practice we perform the convolutions in the time domain, which allows an efficient treatment of the high-frequency components.
For the $G_0 W_0$ scheme only one pass through steps 1--5 is performed. 

When the interaction is considered as spin-dependent, the equation for $W$ in step 4 should be read as a matrix equation in spin-space with a diagonal polarization $\Pi$ and an off-diagonal bare interaction $v_k = U \sigma_x$. Its solution for the diagonal screened interaction yields
\begin{equation}
W_{\sigma\sigma}(\vec{k},i \nu_m) = \frac{U^2 \Pi(\vec{k},i \nu_m)}{1 - [U \Pi(\vec{k},i \nu_m)]^2} \,.
\end{equation}
Additionally, the factor of two in the definition of the polarization, coming from the sum over spins, is dropped in the spin-dependent case.

The computational steps for the self-consistent TMA calculation are as follows:
\begin{enumerate}
\item Initialize the self-energy $\Sigma_\text{TMA}(\vec{k},i\omega_n) = 0$.
\item Calculate the Green's function $G(\vec{k},i\omega_n) = 1 / [i\omega_n + \mu - \epsilon_{\vec{k}} - \Sigma_\text{TMA}(\vec{k},i\omega_n)]$.
\item Calculate the particle-particle polarization function $\Pi_{\text{TMA}}(\vec{k},i\nu_m) = (T/N_k) \sum_{\vec{q}}\sum_{i\omega_n} G(\vec{q},i\omega_n) G(\vec{k}-\vec{q},i\nu_m-i\omega_n)$.
\item Calculate the $T$-matrix $\mathcal{T}(\vec{k},i\nu_m) = -U/[1+U\Pi_{\text{TMA}}(\vec{k},i\nu_m)]$.
\item Calculate the new self-energy $\Sigma_{\text{TMA}}(\vec{k},i\omega_n) = -(T/N_k) \sum_{\vec{q}} \sum_{i\nu_m} \mathcal{T}(\vec{q},i\nu_m) G(\vec{q}-\vec{k},i\nu_m-i\omega_n)$.
\item Go to step 2 until $\Sigma_\text{TMA}(\vec{k},i\omega_n)$ converges.
\end{enumerate}
For the non-self-consistent TMA scheme (TMA$_0$) only one pass through the steps 1--5 is performed.

Finally, the procedures for the self-consistent FLEX calculation are as follows:
\begin{enumerate}
\item Initialize the self-energy $\Sigma_\text{FLEX}(\vec{k},i\omega_n) = 0$.
\item Calculate the Green's function $G(\vec{k},i\omega_n) = 1 / [i\omega_n + \mu - \epsilon_{\vec{k}} - \Sigma_\text{FLEX}(\vec{k},i\omega_n)]$.
\item Calculate the particle-hole polarization function $\Pi_{\text{ph}}(\vec{k},i\nu_m) = (T/N_k) \sum_{\vec{q}} \sum_{i\omega_n} G(\vec{q},i\omega_n) G(\vec{q}-\vec{k},i\omega_n - i\nu_m)$. 
\item Calculate the particle-particle polarization function $\Pi_{\text{pp}}(\vec{k},i\nu_m) = (T/N_k) \sum_{\vec{q}}\sum_{i\omega_n} G(\vec{q},i\omega_n) G(\vec{k}-\vec{q},i\nu_m-i\omega_n)$.
\item Calculate the charge susceptibility $\chi_c(\vec{q},i\nu_m) = \Pi_{\text{ph}}(\vec{q},i\nu_m) / [ 1 - U\Pi_{\text{ph}}(\vec{q},i\nu_m) ]$.
\item Calculate the spin susceptibility $\chi_s(\vec{q},i\nu_m) = \Pi_{\text{ph}}(\vec{q},i\nu_m) / [1 + U\Pi_{\text{ph}}(\vec{q},i\nu_m)]$.
\item Calculate the effective interaction for the particle-hole channel $V_{\text{ph}}(\vec{q},i\nu_m) = U^2 \left[ \frac{3}{2} \chi_s(\vec{q},i\nu_m) + \frac{1}{2} \chi_c(\vec{q},i\nu_m) + \Pi_{\text{ph}}(\vec{q},i\nu_m)  \right]$.
\item Calculate the effective interaction for the particle-particle channel $V_{\text{pp}}(\vec{q},i\nu_m) = U / [1+U \Pi_{\text{pp}}(\vec{q},i\nu_m)] + U^2 \Pi_{\text{pp}}(\vec{q},i\nu_m)$.
\item Calculate the new self-energy $\Sigma_{\text{FLEX}}(\vec{k},i\omega_n) = (T/N_k) \sum_{\vec{q}}\sum_{i\nu_m} [V_{\text{ph}}(\vec{q},i\nu_m) G(\vec{k}-\vec{q},i\omega_n - i\nu_m) + V_{\text{pp}}(\vec{q},i\nu_m) G(\vec{q}-\vec{k},i\nu_m-i\omega_n)]$.
\item Go to step 2 until $\Sigma_\text{FLEX}(\vec{k},i\omega_n)$ converges.
\end{enumerate}
Our definitions differ from the original literature\cite{Bickers1989flex1} in the sign of the particle-hole polarization function $\Pi_{\text{ph}}$, where we use the same convention as in the $GW$ scheme, and in our inclusion of the Hartree term in the particle-particle interaction $V_{\text{pp}}$, which corresponds to the $T$-matrix up to the correction for the second-order term included in $V_{\text{ph}}$. For the non-self-consistent FLEX scheme (FLEX$_0$) only one pass through the steps 1--9 is performed.

Note that in all the above calculations the chemical potential $\mu$ has to be adjusted self-consistently to 
ensure convergence at the desired density. In our calculations the $\vec{k}$-summations are discretized in the first Brillouin zone on an equidistant $80 \times 80$ grid. Furthermore, we include the Hartree term in the chemical potential rather than the self-energy. In other words we redefine the chemical potential and self-energy as
\begin{align} \label{eq:mushift}
    \mu' =& \mu - U \bar{n} / 2, & \Sigma' =& \Sigma - U \bar{n} / 2 ,
\end{align}
with $\bar{n} = \langle n_{i \uparrow} + n_{i \downarrow} \rangle$ the number of electrons per site, and start all calculations with a ``bare'' propagator
\begin{align}
    G_0(\vec{k},i \omega_n) =& 1/[i \omega_n + \mu' - \epsilon_{\vec{k}}]
\end{align}
which includes the mean-field effects of the interaction. This choice is mostly relevant for one-shot calculations and corresponds to the practice in \emph{ab initio} $GW$ calculations, which commonly start from a Hartree-Fock or density functional solution.\cite{aulbur1999,Aryasetiawan1998}

\subsection{Diagrammatic Monte Carlo} \label{diagmc}

\begin{figure}
    \includegraphics[width=\columnwidth]{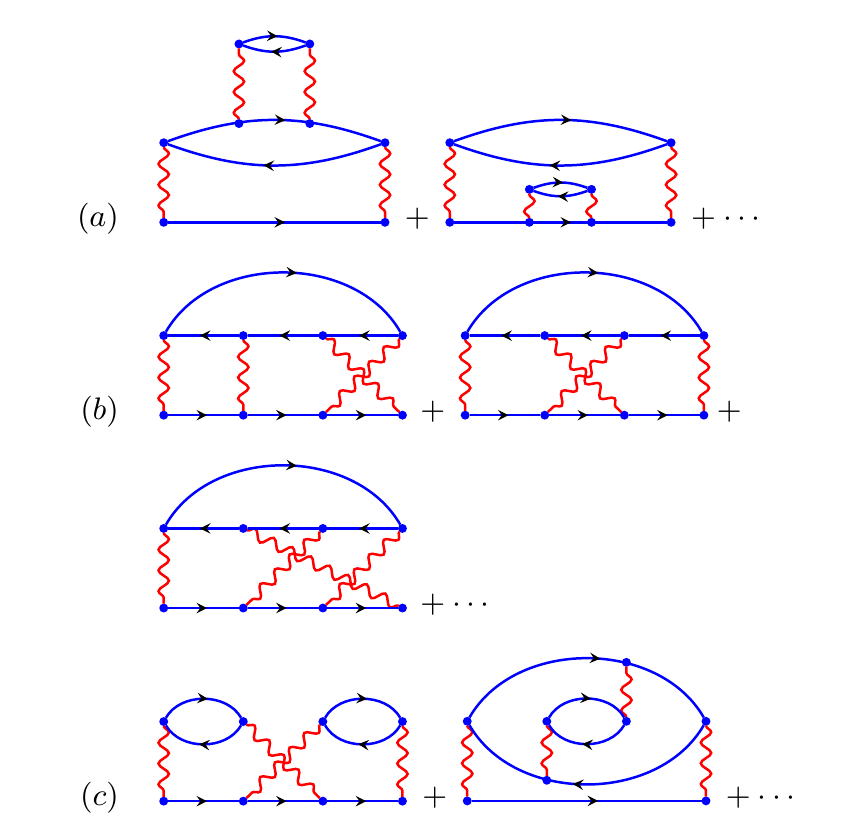}
    \caption{(Color online) Examples of diagram topologies not contained in any of the 
    expansions shown in Fig.~\ref{fig:mbpt}. 
    (a) Self-energy insertions on internal propagator lines. [These are accounted for in self-consistent schemes which use an expansion in terms of the interacting propagator.]
    (b) Ladders with crossed rungs.
    (c) Topologies with more complex vertex corrections. \label{fig:mcdiags}}
\end{figure}

The DiagMC technique\cite{prokofev2007bdm,VanHoucke2010,kozik2010} evaluates a weak-coupling expansion for the self-energy $\Sigma(\vec{k},i\omega_n)$ up to relatively high orders by means of stochastic sampling. In contrast to the approximate schemes discussed above, all diagram topologies are included. A few examples of diagrams neglected in the previous schemes are shown in Fig.~\ref{fig:mcdiags}. While at least FLEX includes all topologies occurring up to third order, the majority of fourth-order diagrams is already neglected. For higher orders, only an exponentially small fraction of the diagrams at a given order is included in approximate methods such as $GW$, TMA or FLEX.

Both the sums over diagram orders and topologies, and the integrals over internal variables are sampled using a Monte Carlo procedure. By restricting the sampling process to one-particle irreducible diagrams the self-energy is computed directly and can then be inserted into Dyson's equation to obtain a single-particle propagator $G(\vec{k},i\omega_n)$ corresponding to an infinite number of diagrams, composed of arbitrary combinations of the explicitly sampled self-energy diagrams. The only systematic error consists in a cutoff of the diagrammatic series at order $N_*$, i.e., the weak-coupling diagrams are generated for orders $N \leq N_*$. Such a cutoff must be introduced because the average sign in the Monte Carlo sampling vanishes exponentially with growing diagram order. By varying $N_*$ and monitoring the convergence of the self-energy, the accessible parameter regime can be determined and the errors can be controlled. We use an expansion in terms of bare propagators which is typically found to converge towards the correct solution in the weak-coupling regime $U \lesssim 4 t$, wherever numerically exact benchmarks are available.\cite{kozik2010,Kozik2014}

\begin{figure}[t!]
    \centering
    \includegraphics[width=\columnwidth]{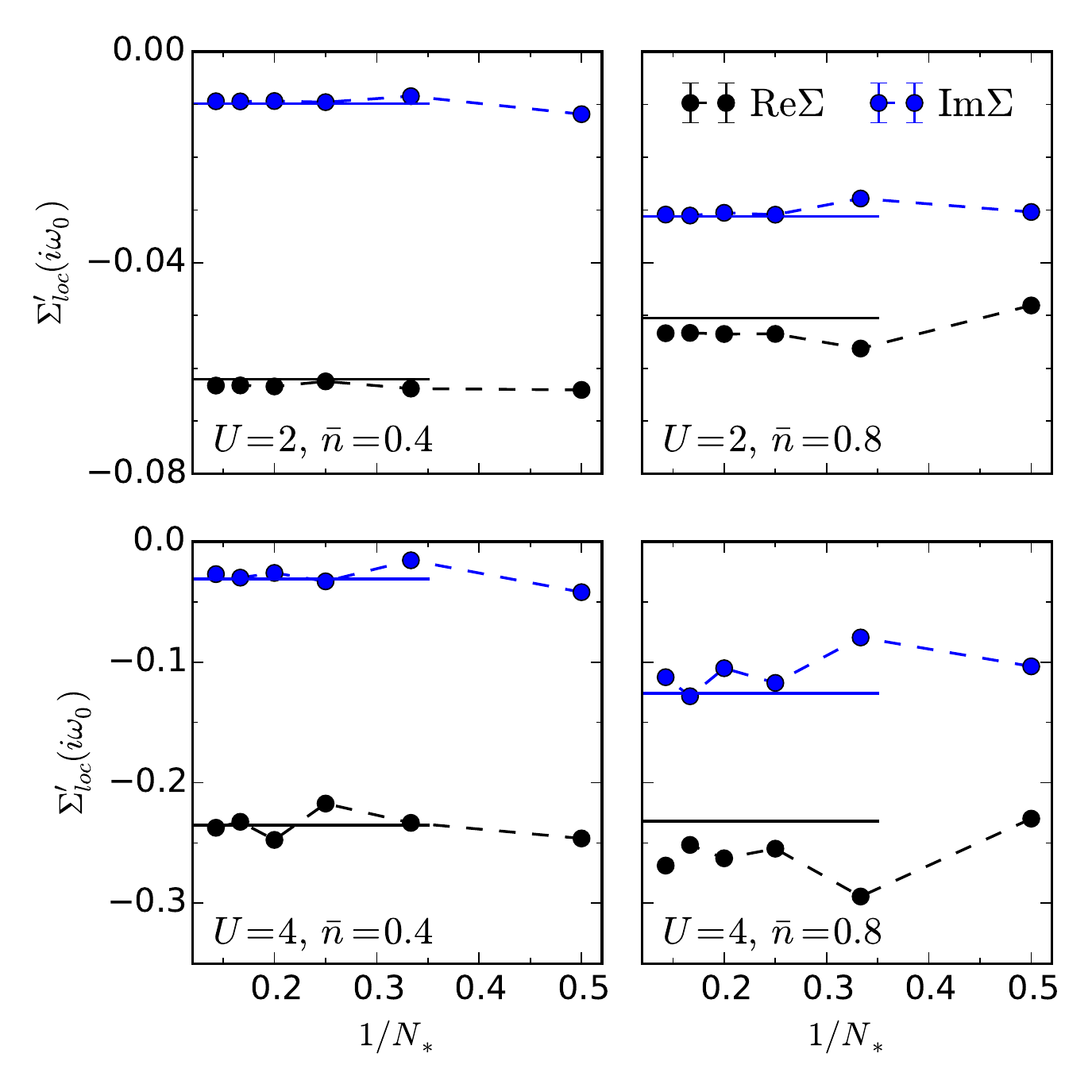}
    \caption{(Color online) Convergence of the weak-coupling series for the local self-energy with diagram order $N_*$ and comparison to the DMFT self-energy (solid horizontal lines). Shown are the real (black) and imaginary (blue) parts at the lowest Matsubara frequency $\omega_0=i \pi T$ for two different fillings $\bar{n}=0.4,0.8$ and two values of the interaction strength $U=2,4$. The temperature is $T=0.1$ in both cases. \label{fig:dmftorder}}
\end{figure}

\begin{figure}
    \centering
    \includegraphics[width=\columnwidth]{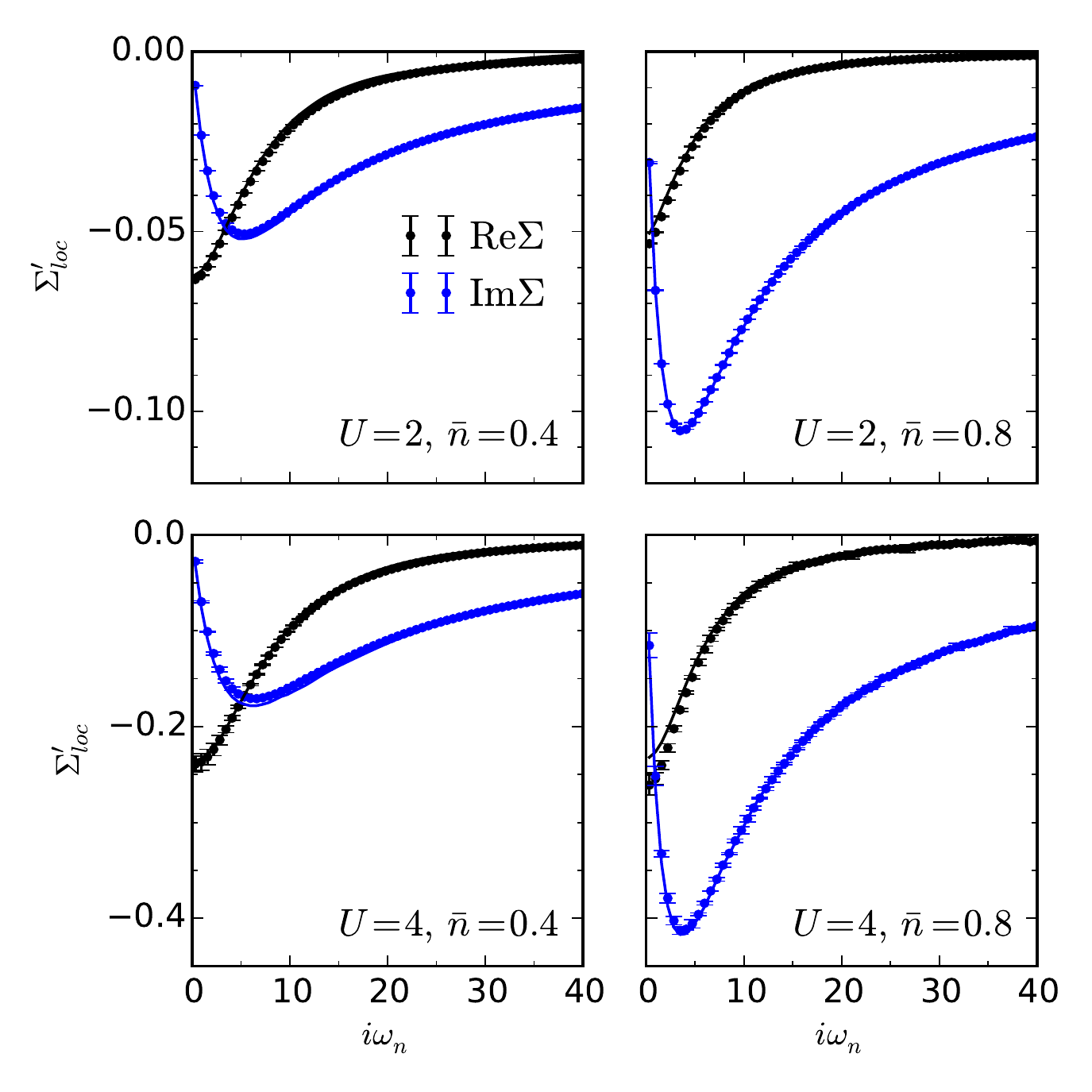}
    \caption{(Color online) Frequency dependence of the local self-energy from DiagMC (black and blue dots for the real and imaginary parts, respectively) and DMFT (solid lines) for the same systems as in Fig.~\ref{fig:dmftorder}. DiagMC error bars cover the results with the three largest cut-off orders $N_*=5,6,7$. \label{fig:dmftfreq}}
\end{figure}

\section{Results}\label{Results}

In the following we compare the self-energies obtained from DMFT, several weak-coupling approximations and MBPT+DMFT schemes to the accurate and well-controlled DiagMC self-energy. We concentrate on the nontrivial part of the self-energy, $\Sigma'$, obtained after subtraction of the Hartree contribution [see Eq.~\eqref{eq:mushift}].

\subsection{Local self-energy}

\begin{figure}[ht!]
    \centering
    \includegraphics[width=\columnwidth]{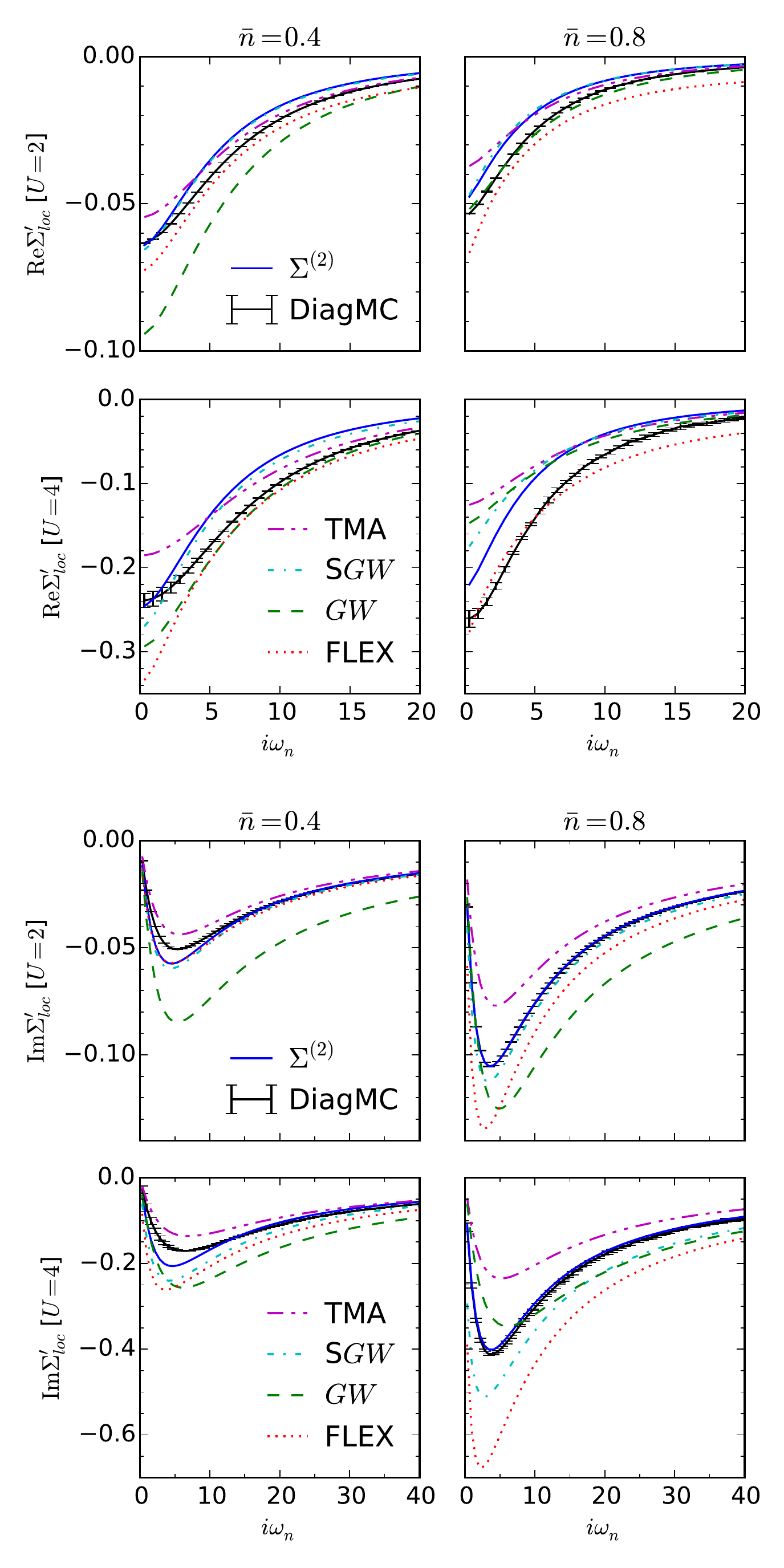}
    \caption{(Color online) Frequency dependence of the local self-energy in the $\Sigma^{(2)}$, $GW$, S$GW$, TMA and FLEX approximation compared to the same DiagMC results as shown in Fig.~\ref{fig:dmftfreq}. The top four panels plot the real parts, and the bottom four panels the imaginary parts of the self-energy. \label{fig:locfreqreal}}
\end{figure}

We start by benchmarking the local self-energy obtained within DMFT. Figure~\ref{fig:dmftorder} plots the lowest Matsubara frequency component of the local self-energy $\Sigma_\text{loc}'(i\omega_0)=\int (\dd k) \, \Sigma'(\vec{k},i\omega_0)$ calculated by the DiagMC method up to order $N_*=7$ and compares it to $\Sigma_\text{DMFT}(i\omega_0)$ for two different site occupancies $\bar{n}=0.4$ and $0.8$ and interaction strengths $U=2$ and $4$. We find that both the real and imaginary parts are quite accurately reproduced by DMFT: $\text{Im}\Sigma_\text{DMFT}(i\omega_0)$ agrees with the DiagMC result within error bars, while $\text{Re}\Sigma_\text{DMFT}(i\omega_0)$ deviates by less than 10\%. 

While the momentum-dependence is neglected, DMFT can capture a nontrivial frequency dependence of the self-energy. Figure~\ref{fig:dmftfreq} shows the comparison of this frequency dependence to the DiagMC results for the same parameter sets. We see that DMFT also predicts the correct frequency dependence of the local self-energy, with maximum relative deviations of less than 10\%.  

We next consider the local component of the self-energy obtained from different weak-coupling approximations. Figure~\ref{fig:locfreqreal} shows the comparison of the $\Sigma^{(2)}$, $GW$, S$GW$, TMA and FLEX results to DiagMC, for the same parameters $U=2,$ $4$ and $\bar{n}=0.4$, $0.8$. While none of the weak-coupling approximations are as accurate as DMFT, the $\Sigma^{(2)}$ (and to a lesser extent the S$GW$) approximation reproduces the exact results rather well. FLEX gives reasonable estimates for the real part but can significantly overestimate the imaginary part, especially near half filling. $GW$ and the TMA yield poor estimates of either the real or imaginary part. 
Not surprisingly, the quality of the TMA decreases with increasing interaction strength and 
away from the dilute limit. The $GW$ approximation, on the other hand, tends to strongly overestimate the self-energy for \emph{weak} interactions. 
Based on these results, we must conclude that schemes involving partial summations of diagrams are less reliable than the simple $\Sigma^{(2)}$ approximation. 
Additionally, $\Sigma^{(2)}$ is the only weak-coupling approximation that correctly captures the exact asymptotic behavior of the self-energy at high frequencies, $\Sigma_{\text{loc}}'(i \omega_n) = U^2 \bar{n}_{\sigma}(1-\bar{n}_{\sigma})/(i \omega_n) + \mathcal{O}(1/(i \omega_n)^2)$, whereas all the other schemes significantly over- or underestimate the coefficient of this tail (not shown).\footnote{Here it should be noted that the band filling $\bar{n}_\sigma$ is uniquely defined only in the self-consistent versions of MBPT. Specifically, in the one-shot application of $\Sigma^{(2)}$ the asymptotic high-frequency behavior is determined by the unperturbed (Hartree-Fock) density $\bar{n}^0_\sigma = G_{0,\text{loc}}(\tau=0^-)$. In general, this differs from the density of the interacting system, $\bar{n}_\sigma = G_{\text{loc}}(\tau=0^-)$, which we fix to the target filling.} %\cite{footnote_highfreq}

The poor performance of the spin-independent $GW$ approximation may at first seem surprising given its widespread and successful use in electronic structure calculations. However, as discussed above, all the other schemes assume a spin-dependent interaction and contain the correct first- and second-order terms of the weak-coupling expansion for a local interaction, whereas $GW$ overestimates the latter term by a factor of two. In \emph{ab initio} calculations of weakly correlated materials, where $GW$ is primarily used, local interactions, and hence violations of the exclusion principle, may be expected to be less relevant. With a non-local interaction, also the other schemes would need to be supplemented by additional exchange diagrams in order to correctly capture all weak-coupling contributions.

Here and in the following we concentrate on the self-consistent versions of MBPT -- except for the simple $\Sigma^{(2)}$ approximation for which we show the one-shot result. In the parameter regime considered here, the difference between one-shot and self-consistent calculations is small for $\bar{n}=0.4$, while there can be significant differences for $\bar{n}=0.8$.  An explicit comparison between one-shot and self-consistent results for the data of Fig.~\ref{fig:locfreqreal} is shown in Appendix~\ref{sec:scvsoneshot}. 

In view of these results, the idea of replacing the local component of the MBPT self-energy by the more reliable DMFT self-energy appears to be reasonable. But before we investigate how this replacement affects different self-consistent schemes, we take a look at the nonlocal components of the self-energy.

\subsection{Nonlocal self-energy}

\begin{figure}[t!]
    \centering
    \includegraphics[width=\columnwidth]{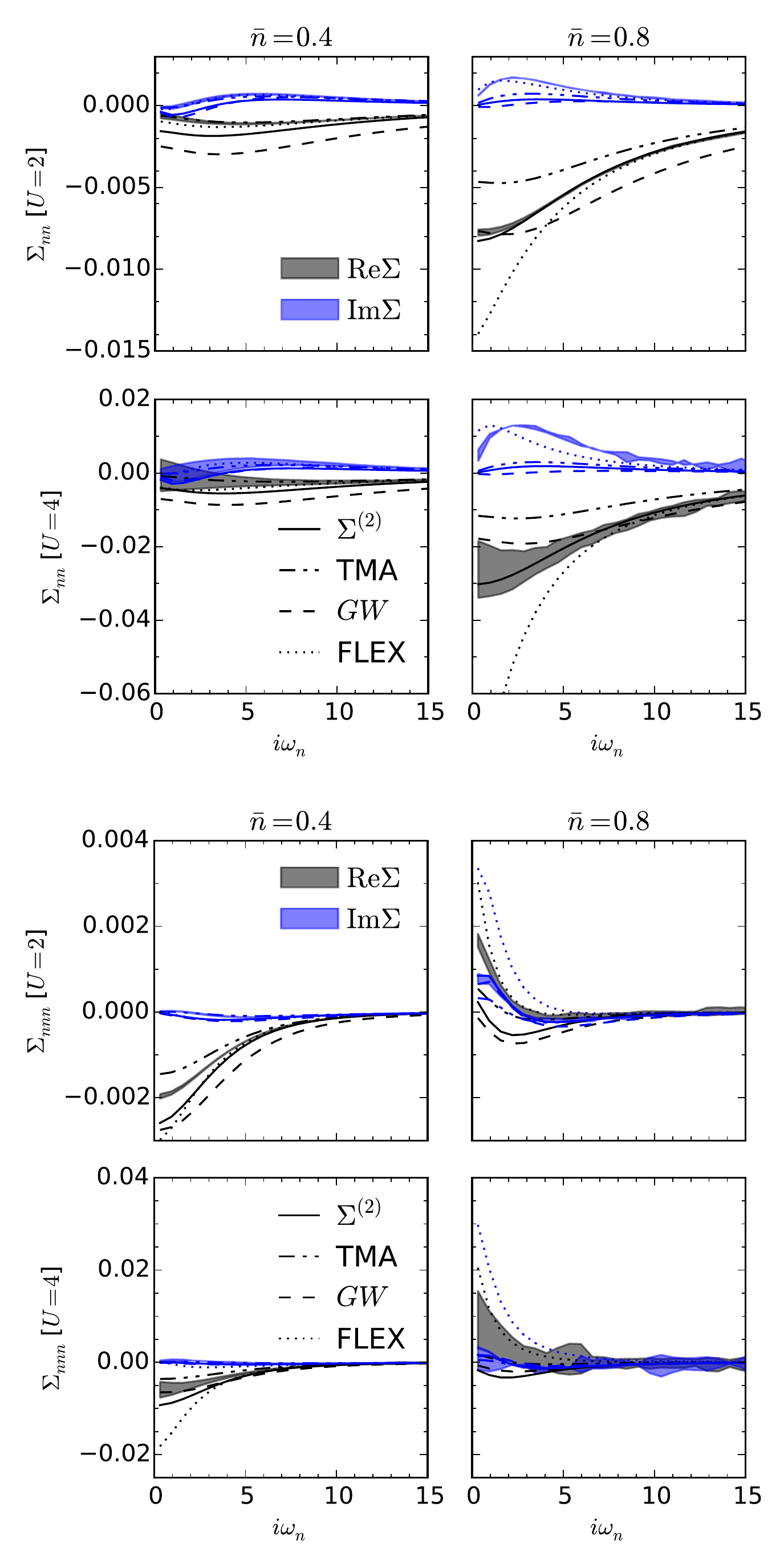}
    \caption{(Color online) Frequency dependence of the nonlocal self-energy for nearest- and next-nearest-neighbor sites in the $\Sigma^{(2)}$, $GW$, TMA and FLEX approximation compared to  DiagMC results. The shaded bands cover stochastic $1 \sigma$ errors around the DiagMC results with the four largest cutoff orders $N_*=4,\dots,7$. The top four panels plot the self-energy for nearest neighbors $\Delta \vec{r}=(1,0)$ and the bottom four panels for next-nearest neighbors along the diagonal of the square lattice $\Delta \vec{r}=(1,1)$. \label{fig:nonlocfreq}}
\end{figure}

Since in the weak-coupling regime considered here the nonlocal self-energy $\Sigma_{ij}$ decays rapidly with the distance $|i-j|$, and it is computationally expensive to obtain DiagMC data with small error bars, we restrict the tests of the nonlocal components to the nearest-neighbor contribution $\Sigma_{nn}$ and the next-nearest-neighbor contribution $\Sigma_{nnn}$. Figure~\ref{fig:nonlocfreq} shows the frequency dependence of these components, again for $U=2$, $4$ and $\bar{n}=0.4$, $0.8$. Both the real (black) and imaginary (blue) parts are plotted in the same panel, and the error estimates of the DiagMC results are indicated by grey and blue shading. We have estimated the systematic uncertainty on the DiagMC data by considering the results for the four largest cutoffs, while the stochastic uncertainty is estimated from 64 independent runs. 

By comparing the $y$-axis scales in Fig.~\ref{fig:nonlocfreq} to the corresponding plots for the local component of the self-energy (Fig.~\ref{fig:locfreqreal}) we see that the $\Sigma_{nn}$ and $\Sigma_{nnn}$ are at least a factor of ten smaller. While the weak-coupling approximations produce nonlocal components of the correct order of magnitude, the relative errors are large. None of the weak-coupling approximations gives reliable results for both the real and imaginary parts. While FLEX seems to work well for the imaginary part of $\Sigma_{nn}$, it gives poor results for the real part and for $\Sigma_{nnn}$. $GW$ and the TMA do not produce very inaccurate results but they are not systematically better than $\Sigma^{(2)}$. To avoid overcrowding the figure, we have not plotted the S$GW$ results, which are typically between those of $GW$ and FLEX. As for the local self-energy, we conclude that there seem to be no obvious benefits from partially summing diagrams beyond the second order.  

\subsection{Combinations of DMFT with weak-coupling approximations}

Since the DMFT approximation provides a good description for the dominant local part of the self-energy, and weak-coupling perturbation theories produce at least a reasonable estimate of the nonlocal components, it is tempting to combine the two approaches. Indeed, such methods have been proposed many years ago, in particular the combination of $\Sigma^{(2)}$ and DMFT \cite{Sun2002} and the combination of $GW$ and DMFT.\cite{Biermann2003} These methods have been designed in particular to treat models with long-ranged Coulomb interactions, based on an extended DMFT (EDMFT) formalism,\cite{Sengupta1995, Si1996, Sun2002, Ayral2013} and because of recent methodological advances related to impurity problems with dynamically screened interactions,\cite{Werner2007, Werner2010} there has been a revival of interest in these approaches.\cite{Ayral2012, Ayral2013, Huang2014} The same techniques can also be applied to model~(\ref{hamilt}) with only an on-site Hubbard repulsion. We will consider here the $\Sigma^{(2)}$+DMFT, $GW$+DMFT and FLEX+DMFT schemes, in which the lattice self-energy is approximated as
\begin{align}
%&\Sigma^\text{$\Sigma^{(2)}$+DMFT}_{ij}(i\omega_n)=\Sigma^\text{DMFT}_{ii}(i\omega_n)\delta_{ij}+\Sigma^{(2)}_{ij}(i\omega_n)(1-\delta_{ij}),\\
%&\Sigma^\text{$GW$+DMFT}_{ij}(i\omega_n)=\Sigma^\text{DMFT}_{ii}(i\omega_n)\delta_{ij}+\Sigma^{GW}_{ij}(i\omega_n)(1-\delta_{ij}),\\
%&\Sigma^\text{FLEX+DMFT}_{ij}(i\omega_n)=\Sigma^\text{DMFT}_{ii}(i\omega_n)\delta_{ij}+\Sigma^\text{FLEX}_{ij}(i\omega_n)(1-\delta_{ij}).
\Sigma^\text{MBPT+DMFT}_{jk}(i\omega_n)=&\Sigma^\text{DMFT}_{jj}(i\omega_n)\delta_{jk}\nonumber\\
&\hspace{5mm}+\Sigma^\text{MBPT}_{jk}(i\omega_n)(1-\delta_{jk}).\label{dc_1}
\end{align}
%
%\textcolor{red}{THIS DOUBLE-COUNTING ACTUALLY REMOVES TOO MANY DIAGRAMS. SHOULD BRIEFLY DISCUSS HERE THE EFFECT OF DIFFERENT DC SCHEMES.}
We have also implemented TMA+DMFT, but will not show these results, because they do not change the main conclusions. 
Note that there are various ways of preventing the double-counting of diagrams. Equation~(\ref{dc_1}) %Here we consider 
corresponds to the simplest approach, i.e., the removal of all the local MBPT self-energy diagrams. This double-counting scheme also removes diagrams with nonlocal propagators, which are not included in the DMFT self-energy. An alternative way of combining the DMFT and MBPT diagrams is
\begin{align}
&\Sigma^\text{MBPT+DMFT}_{jk}(i\omega_n)=\Sigma^\text{DMFT}_{jj}(i\omega_n)\delta_{jk}\nonumber\\
&\hspace{15mm}+\Sigma^\text{MBPT}_{jk}(i\omega_n) - \Sigma^\text{MBPT}_{jj}[G_{jj}](i\omega_n) \delta_{jk},\label{dc_2}
\end{align} 
where $\Sigma^\text{MBPT}_{jj}[G_{jj}](i\omega_n)$ denotes the subset of $\Sigma^\text{MBPT}_{jj}$ diagrams which contains only local propagators $G_{jj}$. We have tested both double counting schemes, but for the parameter sets considered, the differences are rather small. We will show the results for the self-energy (\ref{dc_1}), and comment in the text on the effect of the alternative scheme (\ref{dc_2}), where appropriate. 
 
 \begin{figure}[t!]
    \centering
    \includegraphics[width=\columnwidth]{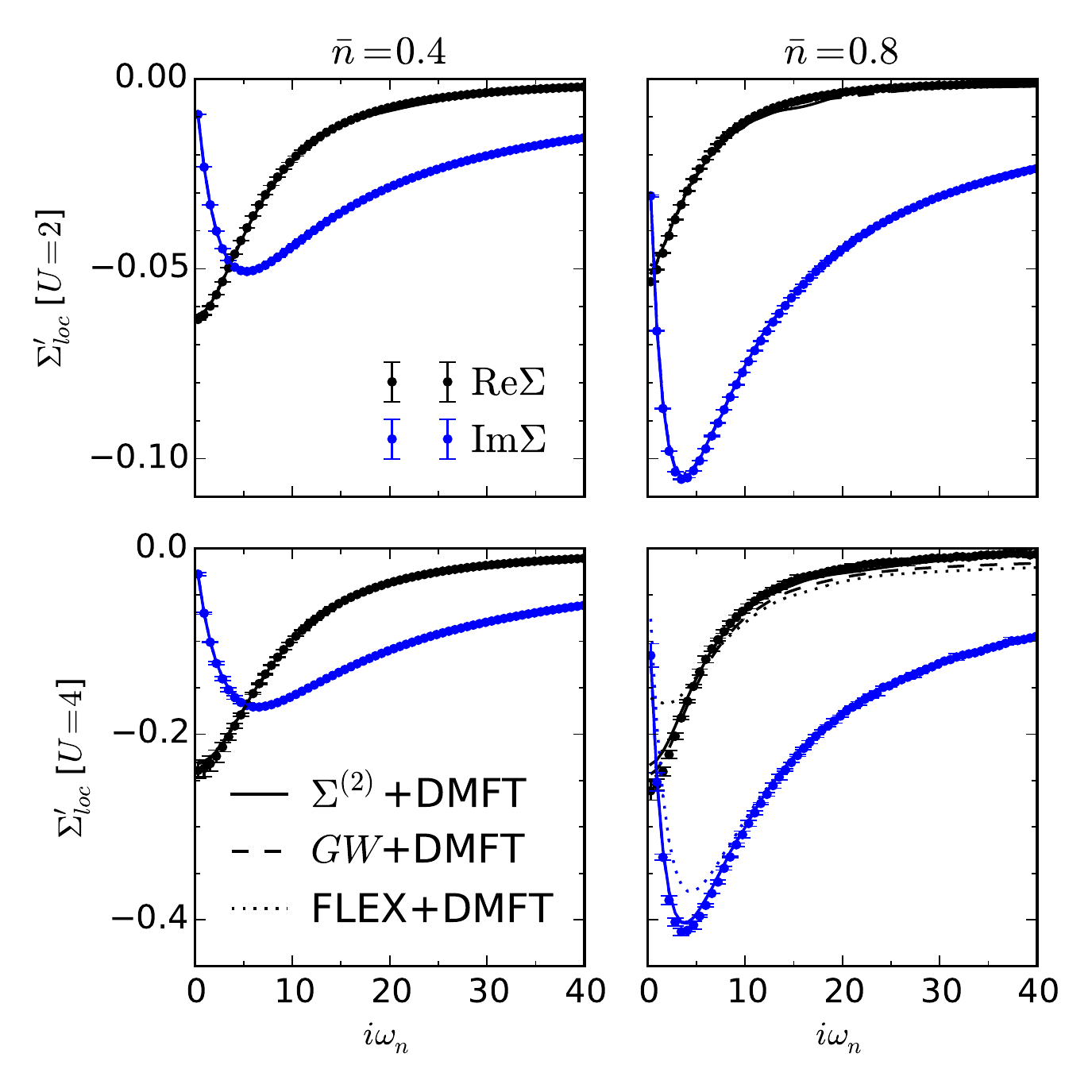}
    \caption{(Color online) Frequency dependence of the local self-energy obtained by $\Sigma^{(2)}$+DMFT (solid lines), $GW$+DMFT (dashed lines), and FLEX+DMFT (dotted lines) compared to the same DiagMC data as in Fig.~\ref{fig:dmftfreq}. \label{fig:gwdmftloc}}
\end{figure}

\begin{figure}[t]
    \centering
    \includegraphics[width=\columnwidth]{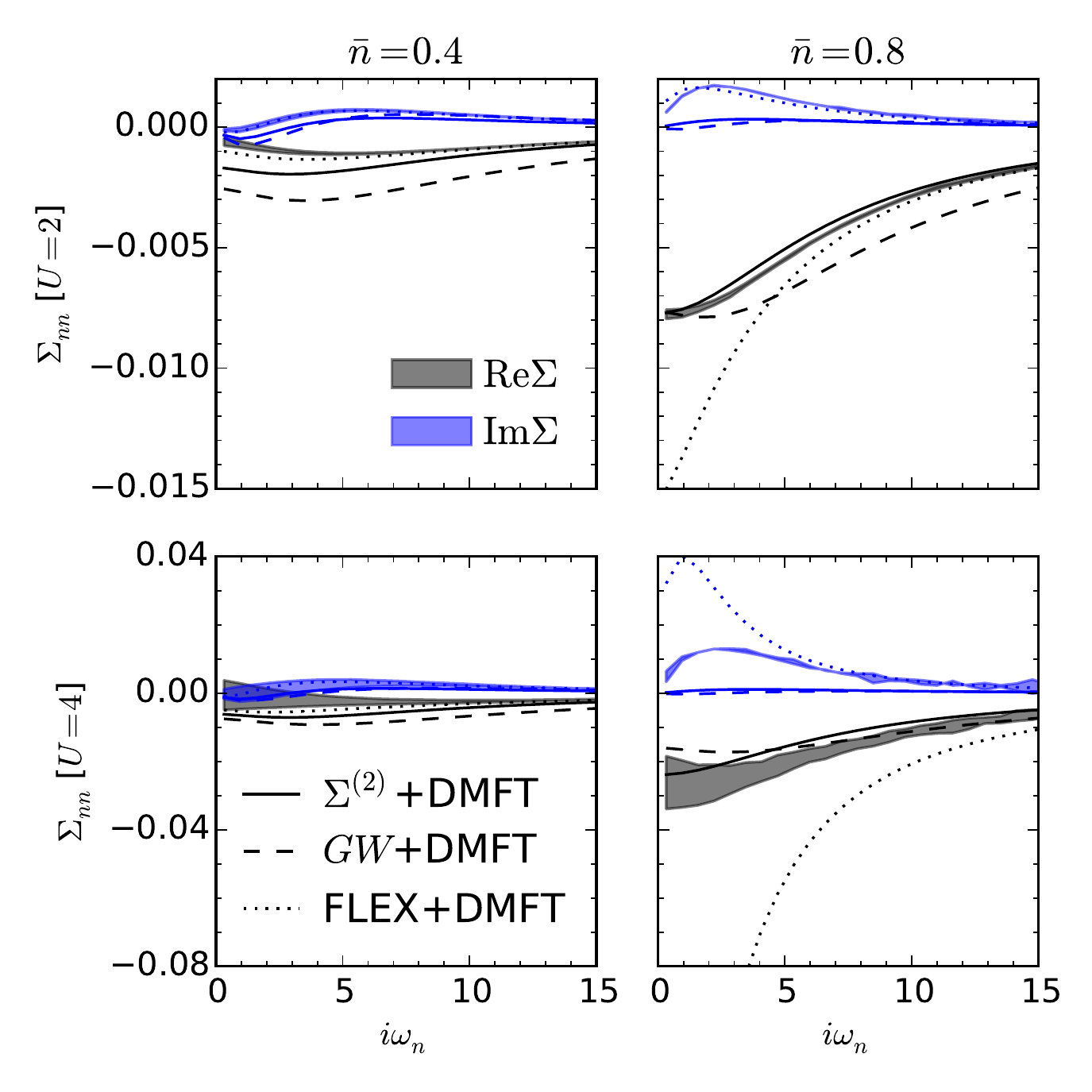}
    \caption{(Color online) Frequency dependence of the nearest-neighbor self-energy obtained by the $\Sigma^{(2)}$+DMFT (solid lines), $GW$+DMFT (dashed lines) and FLEX+DMFT (dotted lines) schemes compared to the same DiagMC data as in the upper half of Fig.~\ref{fig:nonlocfreq}. \label{fig:gwdmftnn}}
\end{figure}

Because the MBPT+DMFT calculations are done self-consistently, it is not easy to identify the subsets of diagrams summed up by these schemes. However, as can be seen in Fig.~\ref{fig:gwdmftloc}, the local $\Sigma$ in the $GW$ + DMFT approximation reproduces the DiagMC result very well. The imaginary part agrees with DiagMC within error bars, and is thus even more accurate than the DMFT result (Fig.~\ref{fig:dmftfreq}), while the accuracy of the real part is comparable to DMFT. Since the real part is very sensitive to the value of the chemical potential, some of these differences may be explained by the uncertainty in the self-consistent calculation of $\mu$. 

\begin{figure*}
    \centering
    \includegraphics[width=2\columnwidth]{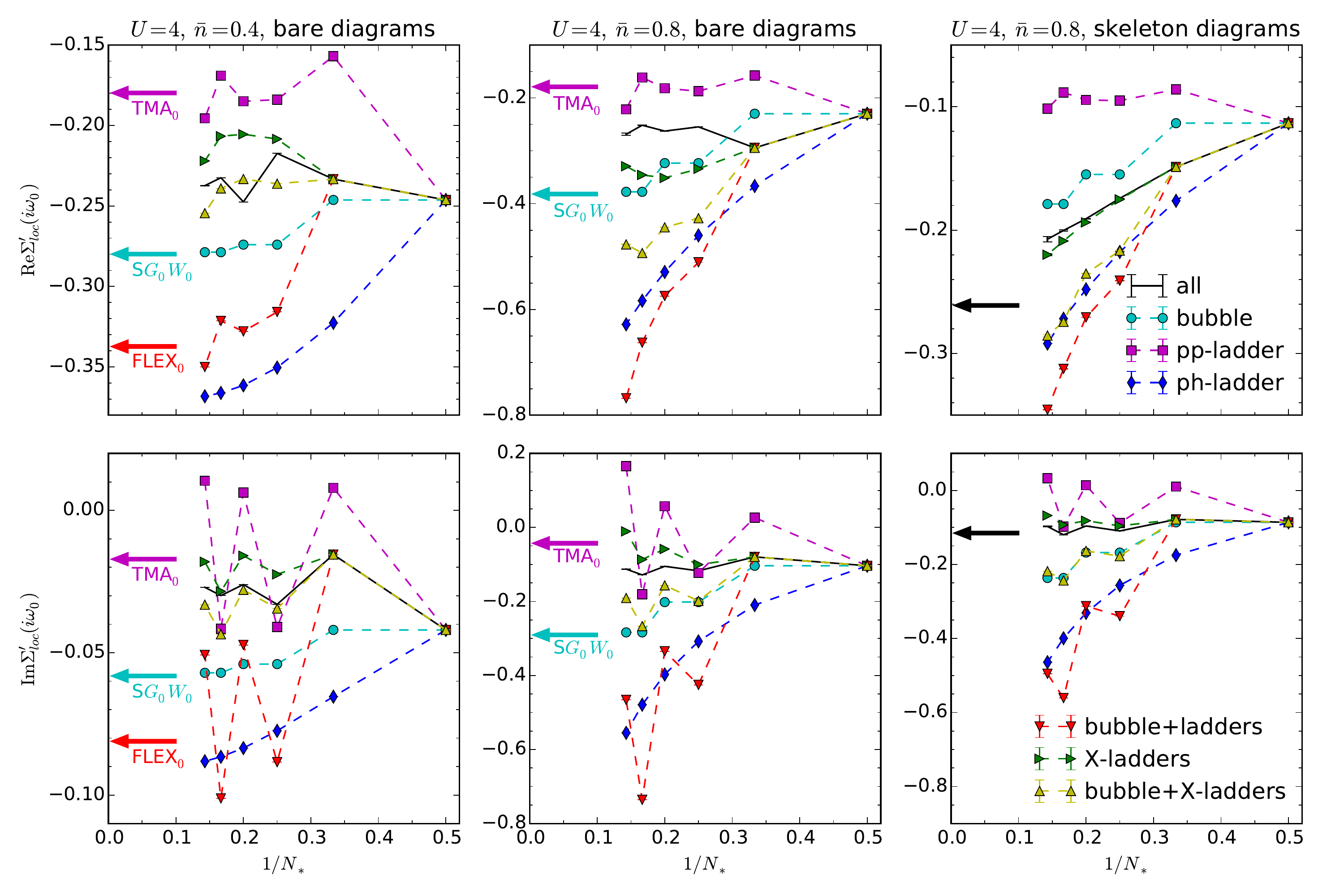}
    \caption{(Color online) Convergence of the local self-energy with diagram order for the full series, sampled in DiagMC, and various subclasses of diagrams. See main text for an explanation of the different diagram classes. The upper (lower) row shows the real (imaginary) part at the lowest Matsubara frequency $i \omega_0=i \pi T$. The left and center columns correspond to the bare series at two different densities, while the right column shows the skeleton series for the same parameters as the central panels. The black arrows in the right panel show the converged DiagMC results, as estimated from the bare series. \label{fig:diagramseries}}
\end{figure*}

In Refs.~\onlinecite{Ayral2013} and \onlinecite{Huang2014} it was found that the combination of $GW$ and EDMFT makes the system more correlated, compared to EDMFT. This conclusion was based on an extended Hubbard model calculation at half filling, with $U=8t$ and nearest neighbor Coulomb repulsion $V\ge0.8t$. Comparing Figs.~\ref{fig:gwdmftloc} and \ref{fig:dmftfreq} we find the opposite effect in the simple Hubbard model away from half filling: the imaginary part of the self-energy is slightly reduced by adding the nonlocal $GW$ self-energy, which means that the system becomes less correlated. This difference may be due to the fact that we consider here a less correlated system, a system away from half filling, or it may indicate that the enhanced correlations in the previous $GW$+EDMFT studies result from a nontrivial interplay between the nonlocal self-energy and nonlocal screening.  In any event, it seems that the addition of the nonlocal $GW$ self-energy can both increase or decrease the local correlations, depending on the parameter regime. 

%\textcolor{red}{NEED A BRIEF DISCUSSION OF THE LOCAL FLEX+DMFT SELFENERGY HERE}
FLEX+DMFT gives improved local self-energies compared to DMFT for $U=2$, and for $U=4$, $\bar{n}=0.4$, but the result for $U=4$, $\bar{n}=0.8$ is significantly less accurate than the DMFT prediction. (With the alternative double counting scheme (\ref{dc_2}), the real part of the self-energy is improved at low Matsubara frequencies, but the imaginary part is overestimated.) Apparently, close to half filling, the feedback from the inaccurate nonlocal FLEX self-energy has a detrimental effect on the local self-energy.  

While the differences to $GW$+DMFT are not very significant, the simple $\Sigma^{(2)}$+DMFT scheme yields the most accurate estimates of the local self-energy, for both interactions and fillings.

%\begin{figure}[t]
%    \centering
%    \includegraphics[width=\columnwidth]{Sigmann_vs_freq_gwdmft}
%    \caption{(Color online) Frequency dependence of the nearest-neighbor self-energy obtained by the $\Sigma^{(2)}$+DMFT (solid lines), $GW$+DMFT (dashed lines) and FLEX+DMFT (dotted lines) schemes compared to the same DiagMC data as in the upper half of Fig.~\ref{fig:nonlocfreq}.
%    }
%    \label{fig:gwdmftnn}
%\end{figure}

Figure~\ref{fig:gwdmftnn} compares the nonlocal self-energy component $\Sigma_{nn}$ obtained from the $\Sigma^{(2)}$+DMFT, $GW$+DMFT and FLEX+DMFT calculations to the DiagMC results. The comparison between the MBPT results and MBPT+DMFT are shown in Appendix~\ref{sec:dmftnloc}. These results illustrate how the self-consistent feedback of the DMFT self-energy into the MBPT scheme affects the nonlocal self-energy. 
%Apparently, the self-consistent feedback of the local DMFT self-energy into the $GW$ calculation results only in small changes of  $\Sigma_{nn}$ compared to the standard $GW$ calculation in the weak-coupling regime considered here and there is no systematic improvement of these nonlocal components. 
%\todo{Discuss difference between X+DMFT methods and move GW vs. GW+DMFT comparison to separate discussion/appendix.}
%Similarly, the nonlocal self-energies from the $\Sigma^{(2)}$+DMFT scheme are slightly less accurate than in the $\Sigma^{(2)}$ approximation. 
%\textcolor{red}{NEED A SENTENCE ABOUT FLEX+DMFT}
In the case of $\Sigma^{(2)}$+DMFT and $GW$+DMFT, the change with respect to the nonlocal $\Sigma^{(2)}$ and $GW$ self-energy is small and there is no systematic improvement of the nonlocal components. For FLEX+DMFT, the conclusion is similar in the case of $U=2$ and $U=4$, $\bar{n}=0.4$, while for $U=4$, $\bar{n}=0.8$ FLEX+DMFT is significantly less accurate than FLEX. (While the double counting scheme (\ref{dc_2}) improves the results somewhat, both the real and imaginary parts of $\Sigma_{nn}$ are still significantly overestimated.) Hence, in the parameter regime where MBPT is not too inaccurate, the local self-energy is apparently improved in the MBPT+DMFT approach, while the nonlocal components of $\Sigma$ are almost unchanged, and do not systematically benefit from the additional local self-energy diagrams in the nonlocal propagators. If the MBPT result deviates strongly from the correct solution, as is the case with FLEX in the intermediate coupling regime close to half filling, then the self-consistent combination with DMFT has detrimental effects on both the local and nonlocal components of the self-energy. 

\subsection{Relevant diagrams}

As discussed in Section~\ref{sec:weak}, a basic assumption underlying approximate schemes such as $GW$ and FLEX is that specific diagram topologies with a rather simple structure contain the relevant physics, at least in certain parameter regimes, such that the summation can be restricted to a tractable subset. In order to test this assumption and possibly identify the relevant subsets, we have implemented a classification scheme for the sampled diagrams in our DiagMC code. This allows us to check, order by order, the respective contributions from $GW$-type bubble diagrams or the particle-particle (pp) and particle-hole (ph) ladders included in the TMA and FLEX approximations. In addition, we consider the class of generalized ladder diagrams (``X-ladders'' for brevity) that includes not only the pp- and ph-ladders but also those with crossed rungs, some examples of which are displayed in Fig.~\ref{fig:mcdiags} (b).

Here, we concentrate on the case $U=4$ and study the evolution of $\Sigma_\text{loc}(i\omega_0)$ with increasing diagram order. We first focus on the bare expansion in terms of the noninteracting propagator $G_0$. The left panels of Fig.~\ref{fig:diagramseries} show data for $\bar{n}=0.4$, with the solid black curve corresponding to the DiagMC result which sums up all diagram topologies. The other curves correspond to the above-mentioned families of diagrams and their combinations. We note that the ``{bubble}'' diagrams correspond to those included in a spin-dependent $G_0 W_0$ calculation and the ``{pp-ladder}'' to a one-shot TMA$_0$ scheme, while the ``{bubble+ladders}'' curves contain exactly the topologies included in a one-shot FLEX calculation. We indicate the results of these one-shot calculations (with the same chemical potential as used in the corresponding DiagMC simulation) with colored arrows on the $y$-axis.

We see that both the bare particle-particle and particle-hole ladders start to deviate significantly from the exact result for orders $\ge 3$, albeit in opposite ways. The bare bubble series seems to be slightly better behaved although it tends to worsen rather than improve the second-order result, in agreement with the findings for the S$G_0W_0$ approximation. While the combination of the particle-particle and particle-hole ladders does not help much, the inclusion of diagrams with crossed rungs in X-ladders does improve the result. This finding is consistent with the intuition of Bickers and White,\cite{Bickers1991} who suggested that ladders with crossed rungs should strongly renormalize the particle-particle and particle-hole ladder contributions, and argued that one should therefore work with a renormalized $U$. (It should be kept in mind that the X-ladder class of diagrams cannot be summed analytically via a Dyson equation.) At least for $\bar{n}=0.4$, the sum of bubbles and X-ladders yields a self-energy which is relatively close to the exact result for the diagram orders considered here. 

The situation gets worse closer to half filling ($\bar{n}=0.8$, see middle panels in Fig.~\ref{fig:diagramseries}). Here, the ``{bubble+X-ladders}'' result deviates strongly from the full series, at least for the real part of the self-energy. Also the other diagram families either converge to wrong values or show no sign of convergence up to the seventh order. This instability is also evident in the FLEX calculations, which need to be initialized with a chemical potential corresponding to a lower filling in order to avoid diverging susceptibilities in the first iteration. Consequently, there are no FLEX$_0$ results indicated in the central panels. Overall, it is clear that none of these families of diagrams yields a systematically better approximation of the local self-energy than the second-order $\Sigma^{(2)}$ contribution. Apparently, the cancellation effects among higher order contributions are so subtle that essentially all diagram topologies must be considered, and the restriction to a subset of ladder or bubble type diagrams cannot be justified. This is further corroborated by the observation that all the subclasses converge, if at all, far less regularly at large orders than the sum of all topologies. Even the X-ladders class, which grows exponentially with diagram order, exhibits seemingly erratic kinks beyond the fifth order, which are apparently canceled by other diagrams, since they are not visible in the sum of all diagrams.

One may wonder whether the situation can be improved by considering only two-particle irreducible skeleton diagrams and replacing the bare propagators $G_0$ by self-consistently computed interacting Green's functions $G$. In order to check this hypothesis we conducted a DiagMC sampling of skeleton diagrams where the propagators are dressed with the self-energy obtained from a previous sampling of the bare series up to sixth order. While such self-consistent calculations sum up more diagrams, the right panels of Fig.~\ref{fig:diagramseries} show that the boldified diagrammatic series converges more slowly than the bare series. This observation is in accord with the recent results of Ref.~\onlinecite{Kozik2014}, where the skeleton series was found to converge very slowly at intermediate interaction $U \sim 4t$. At larger interaction $U \gg 4t$, the skeleton series is even reported to converge to an unphysical solution, whereas the bare expansion shows no such pathological behavior. For the shown parameters the bold X-ladders result is close to DiagMC, but this good agreement appears to be accidental since the corresponding curves at other frequencies significantly deviate from each other, with the X-ladders seemingly converging to incorrect values (not shown).

\section{Conclusions}\label{Conclusions}

We have performed a systematic study of the accuracy of various approximate diagrammatic schemes for the solution of the 2D Hubbard model. By comparing the self-energies obtained from widely used MBPT approaches and DMFT to the well-controlled DiagMC result we were able to assess the quality of the approximations in the weak-coupling regime. We have also measured order by order the contribution of different diagram classes in order to track their convergence properties. The main conclusion is that none of the conventional schemes like $GW$, TMA or FLEX, which sum up bubble and/or ladder diagrams, provides a systematic improvement over the simple $\Sigma^{(2)}$ approximation, and in fact often yield considerably less accurate results. The systematic bias and/or the erratic convergence properties of these schemes with diagram order indicate that the corresponding small subclasses of diagrams do not capture the dominant contributions to the self-energy, and that the corrections from the neglected diagrams are significant. Even by considering additional diagram topologies such as X-ladders, we were not able to identify a `relevant subset' of diagrams. It thus appears that in general, the partial summation of ladder or bubble type diagrams is not a valid approximation, because essentially all diagram topologies are relevant. At least in the weak-coupling regime, stopping at the second order ($\Sigma^{(2)}$) is more reliable than performing uncontrolled summations. Although we cannot access the intermediate and strong-coupling regime with DiagMC, it seems unlikely that a weak-coupling based MBPT approach which is found to be unreliable in the weak-coupling regime can be trusted in the more strongly correlated regime. 
%Our findings thus put a question mark behind the use of $GW$ or FLEX (both the one-shot and self-consistent variants) in studies of lattice models or materials with substantial correlations, such as transition metal, lanthanide or actinide compounds. 
While MBPT methods have been employed by many groups to study transition metal and actinide compounds,\cite{Onari2014,Roedel2009, Aryasetiawan1996, Jiang2009} our findings put a question mark behind the use of $GW$ or FLEX (both the one-shot and self-consistent variants) in studies of lattice models or materials with substantial local correlations.

%On the other hand, for 
For the local part of the self-energy, the DMFT approximation, which is nonperturbative and sums all diagrams made from local propagators, provides a good approximation. This class of diagrams can however not be summed by a simple Dyson-type equation, but requires a self-consistent impurity model calculation. At least in the weak-coupling regime, where the nonlocal components of the self-energy are small, and as we have shown are reasonably described by many-body perturbation approaches such as $\Sigma^{(2)}$ or $GW$, it makes sense to combine the two approaches by adding the nonlocal component of, e.g., the $GW$ self-energy to the local DMFT self-energy. We have tested several MBPT+DMFT schemes and found that for $\Sigma^{(2)}$+DMFT and $GW$+DMFT the feedback from the nonlocal component in the self-consistency loop improves in particular the local self-energy, which becomes very accurate. The nonlocal components are not systematically improved compared to the pure MBPT result, but of comparable accuracy. In FLEX+DMFT, the inaccuracy of the FLEX contribution near half filling can lead to self-energies which are significantly less accurate than the DMFT prediction. 

While $GW$+DMFT has been found to underestimate the $k$-dependence of the self-energy in the intermediate coupling regime,\cite{Ayral2013, Huang2014} compared to cluster DMFT calculations,\cite{Werner2009, Gull2009} this result is not really surprising. The $GW$ method has been primarily designed to capture the effect of screening from long-ranged Coulomb interactions and to avoid divergences in the diagrammatic expansion in terms of the bare Coulomb interaction. This is very important for the proper description of materials,\cite{Werner2012, Werner2014} but does not play a role in the Hubbard model with purely on-site interactions considered in this study. The main target for $GW$+DMFT and related approaches is thus the realistic simulation of (three-dimensional) compounds, where the $k$-dependence can be expected to be small, while the effect of dynamical screening may be significant.   
In this case, the DMFT loop should also involve a self-consistent calculation of the screened interaction\cite{Biermann2003, Ayral2013} and the impurity model must be extended to one with retarded density-density interactions, or even retarded spin-flip terms.

In this work, we have focused on methods which combine DMFT and MBPT at the single-particle (self-energy) level. These should be distinguished from more sophisticated, but also computationally much more expensive methods which attempt such a combination at the two-particle (vertex) level. Since the Coulomb interaction is a two-particle operator, the latter approach may be expected to yield better results and access to the particularly interesting intermediate coupling regime. Systematic tests of methods such as the dynamical vertex approximation or dual fermion based schemes against DiagMC results could provide valuable insights into the virtues and limitations of these vertex based approaches. 

\begin{acknowledgments}
We thank A.~Lichtenstein, T.~Oka, J.~Otsuki and N.~Tsuji for useful discussions. We acknowledge enlightening discussions with E.~Kozik and M.~Troyer during the development of the DiagMC code and thank M.~Troyer for his contributions to writing this paper and E.~Kozik for comments on the manuscript. The calculations have been performed on the M{\"o}nch and Brutus clusters of ETH Zurich and the UniFr cluster. We acknowledge support by the Swiss National Science Foundation (grant No. 200021\_140648) and the European Research Council through ERC Advanced Grant SIMCOFE. The simulations and data evaluation made use of the ALPS libraries.\cite{ALPS2,ALPS13}
\end{acknowledgments}

\appendix

\section{Self-consistent vs.\ one-shot calculations} \label{sec:scvsoneshot}
\begin{figure*}[!ht]
    \centering
    \includegraphics[width=\textwidth]{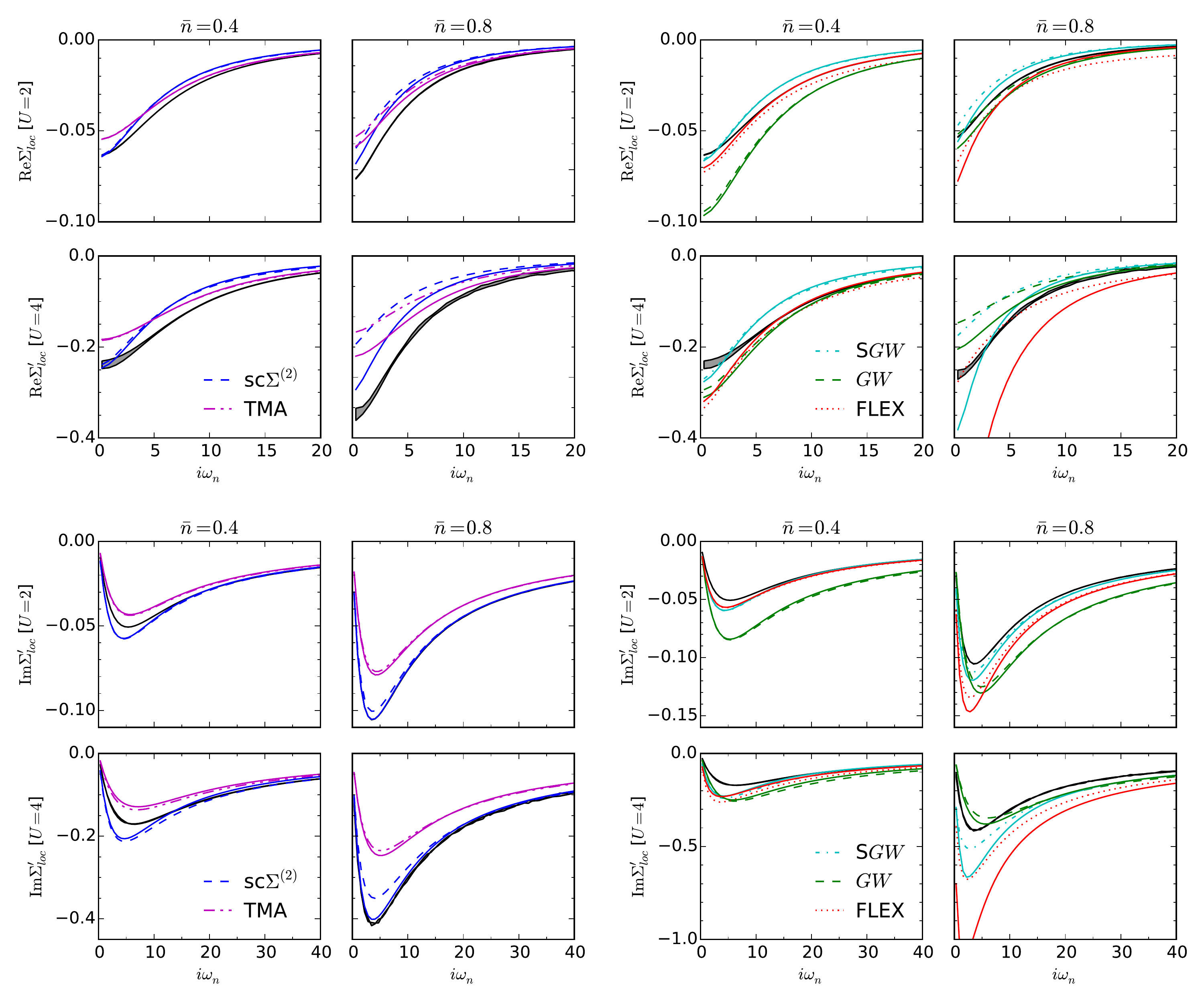}
    \caption{(Color online) Comparison of self-consistent (broken lines) and one-shot (solid lines) results for the real and imaginary parts of the local self-energy from the various weak-coupling approximations.\label{fig:scvsoneshot}}
\end{figure*}

In the parameter regime considered, the differences between one-shot and self-consistent weak-coupling approximations are typically rather small compared to the differences between different diagrammatic approximations. Figure~\ref{fig:scvsoneshot} shows the results for the local self-energy computed using the $\Sigma^{(2)}$, TMA, $GW$, S$GW$ and FLEX approximations for $U=2$, $4$ and $\bar{n}=0.4$, $0.8$. For reference, the respective DiagMC result is indicated by a grey band in each panel. For $\bar{n}=0.4$, the self-consistent resummation changes the self-energy only slightly. For $\bar{n}=0.8$ and weak interaction ($U=2$) the differences between the one-shot and self-consistent schemes are smaller than those between the various weak-coupling approximations. In the vicinity of half filling and for stronger interaction ($U=4$, $\bar{n}=0.8$ in Fig.~\ref{fig:scvsoneshot}), the difference becomes significant. Here in particular the spin-dependent $GW$ and FLEX approximations are close to a pole in the expression for the effective interaction and therefore very sensitive to changes in the polarization. Dressing the propagator reduces the polarization's magnitude and hence moves the expression away from the pole, reducing the resulting self-energy.

\section{Effect of DMFT corrections on the nonlocal self-energy} \label{sec:dmftnloc}
\begin{figure*}[!ht]
    \centering
    \includegraphics[width=\textwidth]{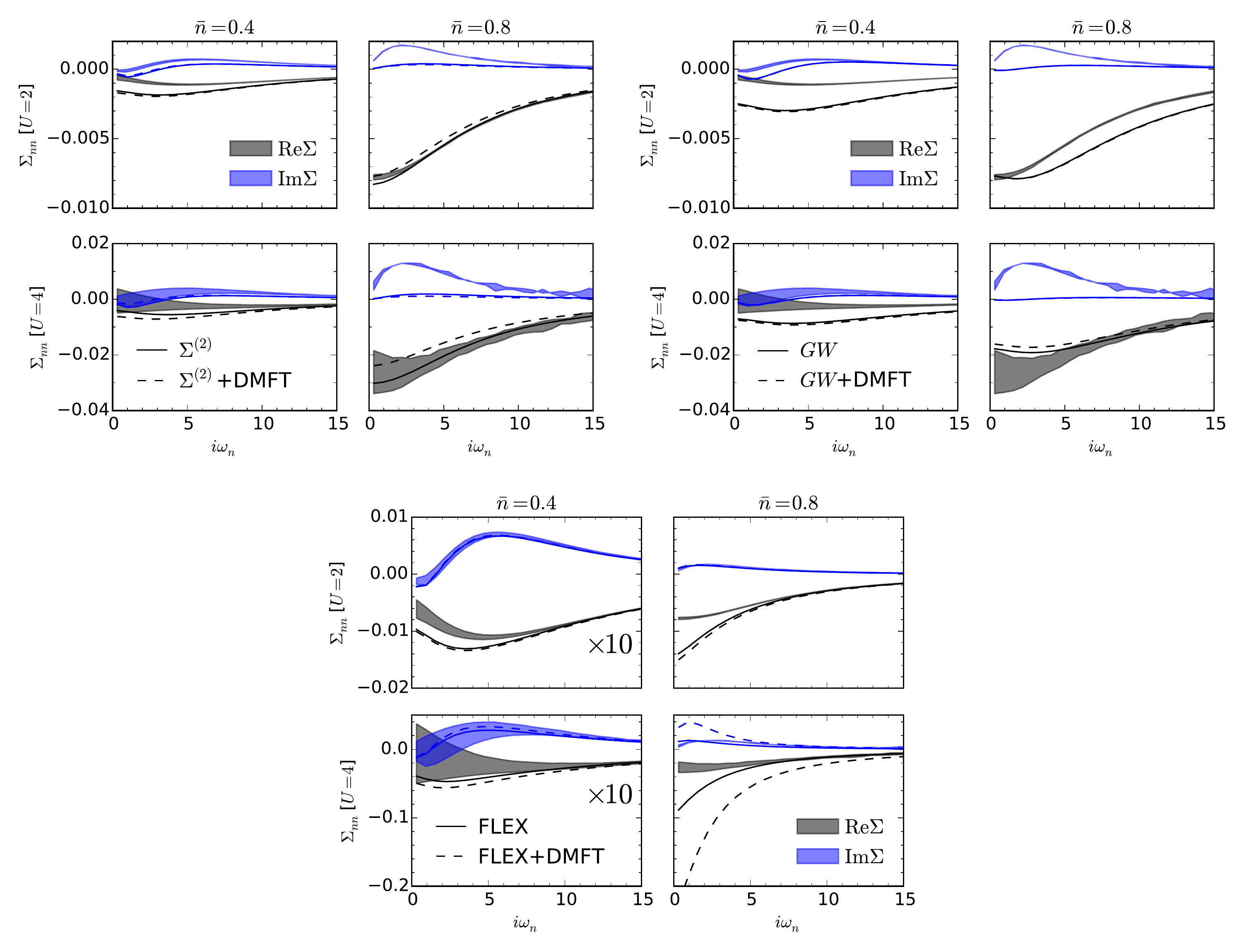}
    \caption{(Color online) Comparison of the self-energy for nearest neighbor sites from $\Sigma^{(2)}$ (top left), $GW$ (top right) and FLEX (bottom) with the results from the corresponding MBPT+DMFT approaches. In the FLEX case, we have multiplied the results for the lower filling $\bar{n}=0.4$ by ten, to make them visible on the same scale as those for $\bar{n}=0.8$. \label{fig:gwvsgwdmft}}
\end{figure*}

Figure~\ref{fig:gwvsgwdmft} compares the nonlocal self-energy obtained from $GW$, $\Sigma^{(2)}$ and FLEX to the corresponding results produced by the combinations of these MBPT methods with DMFT. We see that the inclusion of additional local diagrams in the MBPT propagators has only a moderate effect -- again with the exception of FLEX in the vicinity of half filling and for stronger interactions -- and does not systematically improve the result. While the double counting scheme (\ref{dc_2}) improves the nonlocal self-energy for FLEX+DMFT, the results for both the real and imaginary parts are still significantly larger than for FLEX, and hence less accurate. 

\bibliography{refs}{}

\end{document}